\documentclass{kluwer}
\usepackage{graphicx}
\begin{document}
\begin{article}
\begin{opening}
\title{Obtaining Statistics of Turbulent Velocity from Astrophysical
Spectral Line Data
}            
\author{A. Lazarian\email{lazarian@astro.wisc.edu}}
\institute{University of Wisconsin-Madison, Dept. of Astronomy}


\runningtitle{Obtaining Velocity Statistics}
\runningauthor{Lazarian}

\begin{abstract} 
Turbulence is a crucial component of dynamics of astrophysical
fluids dynamics, including those of ISM, clusters of galaxies and 
circumstellar regions. 
Doppler shifted spectral lines provide a unique source of   
information on turbulent velocities. We discuss Velocity-Channel
Analysis (VCA) and its offspring Velocity Coordinate Spectrum
(VCS) that are based on the analytical description of the
spectral line statistics.
Those techniques are well suited for studies of supersonic
turbulence. We stress that a
 great advantage of VCS is that it does not necessary
require good spatial resolution.
Addressing the studies of mildly supersonic and subsonic
turbulence we discuss the criterion
that allows to determine whether a traditional
tool for such a research, namely, Velocity Centroids are
dominated by density or velocity. We briefly discuss the use
of higher
order correlations as the means to study intermittency of
turbulence. We discuss observational data available
and prospects of the field.
\end{abstract}


\end{opening}

\section{What can be learned from fluctuations of velocity?}

As a rule astrophysical fluids are turbulent and the turbulence 
is magnetized. This ubiquitous turbulence determines the transport
properties of interstellar medium (see 
Elmegreen \& Falgarone 1996, Stutzki 2001,
Cho et al. 2003) and intracluster medium
(Inogamov \& Sunyaev 2003, Sunyaev, Norman \& Bryan 2003,
see review by Lazarian \& Cho 2004a), many properties of Solar and stellar
winds, accretion disks etc. One may say that to understand heat 
conduction, transport and acceleration of cosmic rays, propagation
of electromagnetic radiation in different astrophysical environments it
is absolutely essential to understand the properties of underlying turbulence.
The mysterious processes of star formation (see McKee \& Tan 2002, 
Elmegreen 2002, Pudritz 2001) and interstellar chemistry (see Falgarone 
1999 and
references therein), shattering and
coagulation of dust (see Lazarian \& Yan 2003 and references therein) 
are also intimately related
to properties of magnetized compressible turbulence (see reviews by
Elmegreen \& Scalo 2004).

From the point
of view of fluid mechanics astrophysical turbulence 
is characterized by huge Reynolds numbers, $Re$, which is  
the inverse ratio of the
eddy turnover time of a parcel of gas to the time required for viscous
forces to slow it appreciably. For $Re\gg 100$ we expect gas to be
turbulent and this is exactly what we observe in HI (for HI $Re\sim 10^8$).

Statistical description is a nearly indispensable strategy when
dealing with turbulence. The big advantage of statistical techniques
is that they extract underlying regularities of the flow and reject
incidental details. Kolmogorov description of unmagnetized incompressible
turbulence is a statistical
one. For instance it predicts that the difference in velocities at
different points in turbulent fluid increases on average
with the separation between points as a cubic root of the separation,
i.e. $|\delta v| \sim l^{1/3}$. In terms of direction-averaged
energy spectrum this gives the famous Kolmogorov
scaling $E(k)\sim 4\pi k^2 P({\bf k})\sim k^{5/3}$, where $P({\bf k})$ 
is a {\it 3D} energy spectrum defined as the Fourier transform of the
correlation function of velocity fluctuations $\xi ({\bf r})=\langle  
\delta v({\bf x})\delta v({\bf x}+{\bf r})\rangle$. Note that in
this paper we use $\langle  ...\rangle$ to denote averaging procedure.

The example above shows the advantages of the statistical approach
to turbulence. For instance, the energy spectrum 
$E(k)dk$ characterizes how much
energy resides at the interval of scales $k, k+dk$. At large scales $l$
which correspond to small wavenumbers $k$ ( i.e. $l\sim 1/k$) one expects
to observe features reflecting energy injection. At small scales
one should see the scales corresponding to
sinks of energy. In general, the shape of the spectrum is
determined by a complex process of non-linear energy transfer and
dissipation. For Kolmogorov turbulence the spectrum 
over the inertial range, i.e. the range where neither  energy injection nor
energy dissipation are important, is
characterized by a single power law and therefore self-similar.
Other types of turbulence, i.e. the turbulence of non-linear waves
or  the turbulence of shocks, are characterized by different power
laws and therefore can be distinguished from the Kolmogorov turbulence
of incompressible eddies.    
  
In view of the above it is not surprising that attempts
to obtain spectra of interstellar turbulence have been numerous. In fact they
date as far back as the 1950s
(see von Horner 1951, Munch 1958,
Wilson et al. 1959). However, various directions
of research achieved various degree of success (see 
Kaplan \& Pickelner 1970, a review by Armstrong, Rickett
\& Spangler 1995). 
For instance, studies of turbulence statistics of ionized media 
were more successful
(see Spangler \& Gwinn 1990) and provided the information of
the statistics of plasma density at scales $10^{8}$-$10^{15}$~cm. 
This research profited
a lot from clear understanding of processes of scintillations and scattering
achieved by theorists (see Narayan \& Goodman 
1989). At the same time 
the intrinsic limitations of the scincillations technique
are due to the limited number of sampling directions, and  relevance only to
ionized gas at extremely small scales.
Moreover, these sort of measurements provide only the density statistics, 
which is an indirect measure of turbulence. 

Velocity statistics is much more coveted turbulence measure. 
Although, it is clear that Doppler broadened lines
are affected by turbulence, recovering of velocity statistics was
extremely challenging without an adequate theoretical insight.
Indeed, both velocity
and density contribute to fluctuations of the intensity in the 
Position-Position-Velocity (PPV) space.

\section{What do velocity centroids tell us?}

Let us  consider ``unnormalized'' velocity centroids:
\begin{equation}
S(\mathbf{X})=\int v_{z}\ \rho_{s}(\mathbf{X},v_{z})\ {\rm d}v_{z},
\label{eq:S}
\end{equation}
where $\rho_{s}$ is the density of emitters in the  PPV space\footnote{ Traditionally
used centroids include normalization by the integral of $\rho_s$. This,
however does not substantially improve the statistics, but makes the
analytical treatment very involved (Lazarian \& Esquivel 2003).}.

In the case of emissivity proportional to the first power of density and
provided that the turbulent region is thin for its radiation,
analytical expressions for structure functions\footnote{Expressions for the correlation 
functions are straightforwardly related to those of structure functions (Monin \& Yaglom 1975). 
The statistics of centroids using correlation functions was used in a later paper by 
Levier (2004).}   of centroids, i.e.
$ 
\left\langle \left[S(\mathbf{X_{1}})-S(\mathbf{X_{2}})\right]^{2}\right\rangle $
 were derived in Lazarian \& Esquivel (2003, henceforth LE03). In that paper the following criterion
for centroids to reflect the statistics of velocity\footnote{LE03 showed that the 
solenoidal component of 
velocity spectral tensor can be obtained from observations
using velocity centroids in this regime.} was established:
\begin{equation}
\left\langle \left[S({\bf X_1})-S({\bf X_2})\right]^2 \right\rangle \gg
\langle V^2 \rangle \left\langle \left[I({\bf X_1})-
I({\bf X_2})\right]^2 \right\rangle,
\label{criterion}
\end{equation}
where $I({\bf X})$ is the intensity $I({\bf X})\equiv \int \rho_s dV$ and 
the velocity dispersion $\langle V^2 \rangle$ can
be obtained using the second moment ot the spectral lines:
\begin{equation}
\langle V^2 \rangle\equiv \langle \int_{\bf X} V^2 \rho_s dV\rangle/
\langle \int_{\bf X} \rho_s dV
\rangle .
\label{V2}
\end{equation}

LE03 proposed to subtract the right hand sight of the expression 
(\ref{criterion})
from the left hand sight of (\ref{criterion}) to obtain {\it modified velocity centroids}
that may still reflect velocity statistics 
even when ordinary centroids are dominated
by density contribution. A numerical study in Esquivel \& Lazarian (2004) confirmed that
 the criterion given by eq~(\ref{criterion}) is correct and revealed that for MHD turbulence
simulations it holds for turbulence with Mach number less than 2. This was consistent with
an earlier analysis of a different set of MHD turbulence data by Brunt \& Mac
Low who observed 
that velocity centroids poorly present velocity statistics for high Mach number turbulence.
Esquivel \& Lazarian (2004) studied modified velocity centroids and
the statistical errors
arising from subtracting the corresponding large numbers in order to find the residual fluctuations. They conclude that at the moment velocity centroids can be identified
as a technique to study subsonic turbulence as well as very mildly supersonic turbulence.

\section{How can supersonic turbulence be studied?}

{\bf What do channel maps tell us?}

Power spectra of fluctuations measured within narrow ranges of velocities
have been observed by different authors at various times (see
Green 1993). Those power spectra were guessed to be related
to underlying turbulence, but what exactly those observed spectra
mean was completely unclear.

This problem was addressed in Lazarian \& Pogosyan (2000, henceforth LP00)
who found the relation between the  the spectrum of
intensity fluctuations in channel maps and underlying spectra of velocity and density.
They found that the power index of the intensity fluctuations depends on the
thickness of the velocity channel (see Table~1).

\begin{table*}
\caption{\label{t:lazarian+pogosyan}
A summary of analytical results for channel map statistics derived in LP00.}
\begin{tabular}{lcc}
\noalign{\smallskip} \hline \noalign{\smallskip}
Slice & Shallow 3-D density & Steep 3-D density\\
thickness & $P_{n} \propto k^{n}$, $n>-3$&$P_{n} \propto k^{n}$, $n<-3$\\
\noalign{\smallskip} \hline \noalign{\smallskip}
2-D intensity spectrum for thin~~slice &
$\propto K^{n+m/2}$    & $\propto
K^{-3+m/2}$   \\
2-D intensity spectrum for thick~~slice & $\propto K^{n}$
& $\propto K^{-3-m/2}$  \\
2-D intensity spectrum for very thick~~slice & $\propto K^{n}$ & $\propto \
K^{n}$  \\
\noalign{\smallskip} \hline \noalign{\smallskip}
\end{tabular}
{{\it thin} means that the channel width $<$ velocity dispersion at the scale under
study}\\
{{\it thick} means that the 
channel width $>$ velocity dispersion at the scale under
study}\\
{{\it very thick} means that a
substantial part of the velocity profile is integrated over}\\
{$m$ is the power-law index of velocity structure
function, i.e. $\langle (v(x+r)-v(x))^2\rangle \sim r^{m}$.}\\
{$K$ is a 2D wavevector in the plane of the slice, $k$ a 3D wavevector.}
\end{table*}

It is easy to see that both steep and shallow underlying density
the power law index
{\it steepens} with the increase of velocity slice
thickness. In the thickest velocity slices the velocity information
is averaged out and it is natural that we get the
density spectral index $n$. The velocity fluctuations dominate in
thin slices, 
and the index $m$ that characterizes the velocity  fluctuations,
i.e. $|\delta v|\sim l^{m/2}$,
can be obtained using thin velocity slices (see Table~1). Note, that the
notion of thin and thick slices depends on a turbulence scale under
study and the same slice can be thick for small scale turbulent fluctuations
and thin for large scale ones. The formal criterion for the slice to be
thick is that {\it the dispersion of turbulent velocities on the scale studied
should be less than the velocity slice thickness}.  Otherwise
the slice is {\it thin}.

One may notice that the spectrum
of intensity in a thin slice gets shallower as the underlying
velocity get steeper. To understand this effect let usconsider turbulence
in incompressible optically thin media. The intensity of emission
in a slice of data is proportional to the number of atoms per
the velocity interval given by the thickness of the data slice.
Thin slice means that the velocity dispersion at the scale of
study is larger than the thickness of a slice. The increase
of the velocity dispersion at a particular scales means that
less and less energy is being emitted within the velocity
interval that defines the slice. As the result the image of
the eddy gets fainter. In other words, the larger is the
dispersion at the scale of the study the less intensity
is registered at this scale within the thin slice of spectral
data. This means that steep velocity spectra that correspond
to the flow with more energy at large scales should produce
intensity distribution within thin slice for which the
more brightness will be at small scales. This is exactly
what our formulae predict for thin slices (see also LP00).

The result above gets obvious when one recalls that the largest
intensities within thin slices are expected from the regions that
are the least perturbed by velocities. If density variations are
also present they modify the result. When the amplitude of density
perturbation becomes larger than the mean density, both the
statistics of density and velocity are imprinted in thin slices.
For small scale asymptotics of thin slices this happens, however, only when the density spectrum
is shallow, i.e. dominated by fluctuations at small scales.

{\bf Is spatial resolution necessary to study statistics of velocity?}

Spatial resolution is essential for centroids and channel maps. However,
the velocity fluctuations are also imprinted on the 
fluctuations of intensity along the velocity coordinate direction. 
The corresponding 3D PPV spectra
were derived in LP00. Table~3 in LP00 states that two terms, one
depending only on velocity and the other depending both on velocity
and density, contribute to the spectrum measured along velocity
coordinate. 
for steep density the intermediate scaling therefore is
$k^{2n/m}$ ($n$ is negative), while the small scale assymptotics scales as 
$k^{-6/m}$. If the density is shallow the situation is refersed,
namely, at larger scales $k^{-6/m}$ assymptotics dominates, while
$k^{2n/m}$ assymptotics is present at smaller scales. The transition
from one assymptotics to another depends on the amplitude of density
fluctuations. Therefore both density and velocity statistics can 
be restrored from the observations this way. If the measurements are done
with an instrument of poor spatial resolution these are the
expected scalings. A further study of these
interesting regime is done in Chepurnov \& Lazarian (2004).
The only requirement for the VCS to operate is for the turbulence 
to be
supersonic and for the instrument to have an adequate {\it
spectral} resolution.

\section{What is the effect of absorption?}

The issues of absorption were worrisome for the researchers from the
very start of the research in the field (see Munch 1958). The erroneous
statements about the effects of absorption on the observed turbulence
statistics are widely spread in the literature. For instance, a
fallacy that absorption allows to observe density fluctuations
localized in the thin surface layers of clouds, i.e. 2D turbulence,
exists (see discussion in LP04).

Using transitions that are less affected by absorption, e.g. HI,
allows frequently to avoid the problem. However, it looks foolish
to disregards the wealth of spectroscopic data only because absorption
is present. A study of absorption effects is given
in LP04. There it was found that for sufficiently thin slices
the scalings obtained in the absence of absorption still hold
provided that the absorption on the scales under study is negligible.
A similar criterion is valid for the VCS.
From the practical point of view, absorption imposes an upper limit
on the scales for which the statistics can be recovered.

If integrated intensity of spectral lines is studied in the presence of
absorption non-trivial effects emerge. Indeed, for optically thin
medium the spectral line integration results in intensity reflecting
the density statistics. LP04 showed that this may not
be any more true for lines affected by absorption. Depending on the
spectral index of velocity and density fluctuations the contributions 
from either from 
density or velocity dominate the integrated intensity fluctuations.
When velocity is dominant a very interesting regime for which
intensity fluctuations show universal behavior, i.e. the
power spectrum $P(K)\sim K^{-3}$  emerges. If density is dominant,
the spectral index of intensity fluctuations is the same
as in the case 
an optically thin cloud. Conditions for these regime as well
as for some more interesting intermediate asymptotic regimes
are outlined in LP04.

\section{How can intermittency be studied?}

Velocity and density power spectra do not provide a complete description
of turbulence. Intermittency of turbulence (its variations in time and space)
and its topology in the presence of different phases are
not described by the power spectrum.
Recent numerical research that employed higher order
correlation functions (Muller \& Biskamp 2000,
Cho, Lazarian \& Vishniac 2002b, 2003) showed them to be a promising tool. 
For instance, the distinction between
the old Iroshnikov-Kraichnan and the GS95 model is difficult
to catch using power spectra with a limited inertial range, but is quite
apparent for fourth order statistics. The difference in physical 
consequences of whether the turbulence dissipates in shocks
or in intermittent vortices may be very substantial. Recent
research (see review by Lazarian \& Cho 2004) 
suggests that dissipation of interstellar motions via vortices is
very important. Although
the scaling of vortex intermittency with the Reynolds number $R$
is not clear, it is suggestive that the intermittency
is increasing with $R$. 

Higher order statistics obtained
from observational data 
were reported for observed velocity\footnote{Whether in all cases the
used centroids reflected the actual velocity statitics is not sure because
the criterion by Lazarian \& Esquivel (2003) has not been applied to the
data.} in Falgarone et al.
(1994)  and for density  in Padoan et al. 2003. 
According to Falgarone \& Puget (1995) and
Falgarone et al. (1995),
the intermittency in vorticity distribution
can result in the outbursts of localized dissipation that
make tiny regions within cold diffuse clouds chemically active.
This is an extremely important conclusion that stimulates
more intensive studies of higher moment statistics from
observations.

\section{What are the niches for different techniques?}

Astrophysical fluids demonstrate turbulent motions at very different 
Mach numbers.
The ``old and good'' velocity centroids are shown (LE03, Esquivel \& 
Lazarian 2004)
 to be reliable only at low Mach numbers. This makes some of the 
earlier results 
obtained, e.g. for 
hypersonic motions in HI using velocity centroids (see Meville-Deschenes et
al. 2003),
as well as some results on molecular clouds somewhat questionable.
Observational testing whether the criterion (\ref{criterion}) is
satisfied is essential for a confident use of centroids to study
velocity statistics.

For supersonic turbulence VCA and VCS present the best bet at the 
moment. The analytical description of intensity fluctuations in 
PPV space obtained in LP00 and LP04 allows to reliably separate
velocity and density contributions to the observed fluctuations.
VCS looks to be the most promising tool as it requires only frequency
resolution to be adequate. It opens an avenue for studies of
turbulence in poorly resolved objects, e.g. for extragalactic  research.

Another limitation on the use of the techniques arises from the
density-velocity correlations in the data. LP00 analytically
studied the effect of velocity-density correlations for the
VCA and concluded that even for the maximal possible level 
of correlation, the scaling of intensity fluctuations 
in thin velocity channels is not affected. Further numerical studies
in Lazarian et al. (2001) and Esquivel et al. (2003) confirmed that the
velocity-density correlations present in compressible MHD
flows do not change the statistics obtained with the VCA. Our ongoing
research shows that velocity centroids seem to be more affected
by the velocity-density correlations, but this is not a dominant 
effect for at least for the low Mach number flows for which the
centroids are applicable. Currently we study the effect of correlations
arising from gravity. 

For studies of higher order statistics velocity centroids present
the best bet for the moment. This places limitations on the data
sets to be studied (e.g. the criterion  (\ref{criterion}) should be satisfied),
and stimulates the search for an alternative techniques.

We also would like to stress that the different techniques discussed
are complementary. For instance, it was noted in LE03 that the velocity
centroids are sensitive only to solenoidal motions, while VCA is affected
by both potential and solenoidal motions. Therefore combining the two 
techniques
it should be possible to study the effects of compressibility in the astrophysical flows. In addition, different techniques are affected differently by
gas temperature. This allows to get insight into the temperature distribution
along the line of sight. 

Obtaining the statistics of velocity turbulence may provide sometimes 
unexpected  bonuses. For instance, the damping scale of turbulence can be
determined if the velocity dispersion is known  at the injection scale.
This scale could provide a standard yardstick for finding the distances
to clouds, e.g. to high velocity clouds.

\section{What do observations tell us?}

Application of the VCA to the Galactic data in LP00 and to
Small Magellanic Cloud in Stanimirovic \& Lazarian (2001)
revealed spectra of 3D velocity fluctuations consistent with
the Kolmogorov scaling. LP00 argued that the same scaling
was expected for the magnetized turbulence appealing to
the Goldreich-Shridhar (1994) model. Esquivel et al. (2003)
used simulations of MHD turbulent flows to show that in spite
of the presence of anisotropy caused by magnetic field the
expected scaling of fluctuations is Kolmogorov. Studies by
Cho\& Lazarian (2002, 2003) revealed that the Kolmogorov-type
scaling is also expected in the compressible MHD flows.
These studies support MHD turbulence model for SMC.

Studies of turbulence are more complicated for the inner parts of the Galaxy,
where (a) two distinct regions at different distances from the observer
contribute to the emissivity for a given velocity and (b) effects of
the absorption are important. However, the analysis in Dickey et al. (2001)
showed that some progress may be made even in those unfavorable
circumstances. Dickey et al. (2001) found the steepening
 of the spectral index with the increase of the velocity slice thickness.
They also observed the spectral index for strongly absorbing direction
approached $-3$ in accordance with the conclusions in LP04.

21-cm absorption provides another way of probing turbulence on small
scales. The absorption depends on the density to temperature ratio
$\rho/T$, rather than to $\rho$ as in the case of emission. However,
 in terms of the VCA this change is not important and we still expect to
see emissivity index steepening as velocity slice thickness increases,
provided that velocity effects are present. In view of
this, results of Deshpande et al. (2001), who did not see such steepening,
can be interpreted as the evidence of the viscous suppression of
turbulence on the scales less than 1~pc. The fluctuations in this
case should be due to density and their shallow spectrum $\sim k^{-2.8}$ may
be related to the damped magnetic structures below the viscous
cutoff (Cho, Lazarian \& Vishniac 2002b). 

Studies of velocity statistics using velocity centroids are
numerous (see O'Dell 1986, Miesch \& Bally 1994, Miesch, Scalo \& Bally 1999).
The analysis of observational data in Miesch \& Bally (1994) provides a range
of power-law indexes. Their results obtained with structure
functions if translated into spectra are consistent with
$E(k)=k^{\beta}$, where $\beta=-1.86$ with the standard
deviation of $0.3$. The Kolmogorov index $-5/3$
falls into the range of the measured values. L1228 exhibits
exactly the Kolmogorov index $-1.66$ as the mean value,
while other low mass
star forming regions L1551 and HH83 exhibit indexes close
to those of shocks, i.e. $\sim -2$. The giant molecular cloud regions show
shallow indexes in the range of $-1.9<\beta<-1.3$ (see Miesch et al. 1999).
It worth noting that Miesch \& Bally (1994)
obtained somewhat more shallow indexes that are closer to the
Kolmogorov value using autocorrelation functions. Those may
be closer to the truth as in the presence
of absorption in the center of lines, minimizing the regular velocity
used for individual centroids might make the results
more reliable. Whether the criterion given by (\ref{criterion})
is satisfied for the above data is not clear. If it is not
satisfied, which is quite possible as the Mach numbers are
large for molecular clouds, then the spectra above reflect
the density rather than velocity. If the criterion happen to
be satisfied a more careful study
taking absorption effects into account is advantageous. 

Apart from testing of 
the particular
scaling laws, studies of turbulence statistics
 should identify sources and injection
scales of the turbulence.
 Is turbulence in molecular clouds a part
of a large scale ISM cascade (see Armstrong et al. 1997)? 
How does the share of the energy within
compressible versus incompressible motions vary within the Galactic
disk? There are examples of questions that can be answered in
future.

{\bf Acknowledgments}{ 
A.L. acknowledges  the support of Center for Magnetic Self-Organization
in Laboratory and Astrophysical Plasmas. Fruitful discussions with
Alexey Chepurnov and Dmitry Pogosyan are acknowledged}.

\end{article}
\end{document}